\documentclass[superscriptaddress,aps,pra,onecolumn,showpacs,nofootinbib,longbibliography,notitlepage]{revtex4-1}
\usepackage{etex}
\usepackage{amsmath,amssymb,amsthm}
\usepackage[colorlinks=true,citecolor=blue,urlcolor=blue]{hyperref}
\usepackage[pdftex]{graphicx}
\usepackage{times,txfonts}
\usepackage{braket}
\usepackage{color}
\usepackage{natbib}
\usepackage{amsmath,blkarray}
\usepackage{mathtools}
\usepackage{latexsym}
\usepackage{subfig}
\usepackage{tabularx, booktabs}
\usepackage{graphics,epstopdf}
\usepackage{graphicx}
\usepackage{float}
\usepackage{graphicx}
\usepackage{amsfonts}
\usepackage{color,soul}

\newcommand{\be}{\begin{equation}}
\newcommand{\ee}{\end{equation}}
\newcommand{\ba}{\begin{eqnarray}}
\newcommand{\ea}{\end{eqnarray}}

\begin{document}
\title{Demonstration of quantum delayed-choice experiment on a quantum computer}    

\author{Pranav D. Chandarana}
\email{pranav.chandarana@gmail.com}
\affiliation{Department of Physics, Fergusson College, FC road, Pune-411004, India}

\author{Angela Anna Baiju}
\email{angelbaiju08@gmail.com}
\affiliation{Department of Physics, National Institute of Technology Calicut, Kerala 673 601, India}

\author{Sumit Mukherjee}
\email{mukherjeesumit93@gmail.com}
\affiliation{Department of Physical Sciences, Indian Institute of Science Education and Research Kolkata, Mohanpur 741246, West Bengal, India}

\author{Antariksha Das}
\email{a.das-1@tudelft.nl}
\affiliation{QuTech, Delft University of Technology, Lorentzweg 1, 2628 CJ Delft, The Netherlands}

\author{Narendra N. Hegade}
\email{narendrahegade5@gmail.com}
\affiliation{International Center of Quantum Artificial Intelligence for Science and Technology (QuArtist), and Department of Physics, Shanghai University, 200444 Shanghai, China}%

\author{Prasanta K. Panigrahi}
\email{pprasanta@iiserkol.ac.in}
\affiliation{Department of Physical Sciences, Indian Institute of Science Education and Research Kolkata, Mohanpur 741246, West Bengal, India}

\begin{abstract}
Wave-particle duality of quantum objects is one of the most striking features of quantum physics and has been widely studied in past decades. Developments of quantum technologies enable us to experimentally realize several quantum phenomena. Observation of wave-particle morphing behavior in the context of the quantum delayed-choice experiment (QDCE) is one of them. Adopting the scheme of QDCE, we demonstrate how the coexistence of wave and particle nature emerges as a consequence of the uncertainty in the quantum controlled experimental setup, using a five-qubit cloud-based quantum processor. We also show that an entanglement-assisted scheme of the same reproduces the predictions of quantum mechanics. We put evidence that a local hidden variable theory is incompatible with quantum mechanical predictions by comparing the variation of intensities obtained from our experiment with hidden variable predictions.
\end{abstract} 
\maketitle
\section*{Introduction}
Quantum Mechanics is known for challenging our usual perceptions of nature. Superposition, wave-particle duality, and entanglement, are perhaps the three most ``mysterious" properties of quantum mechanics that don't go hand-in-hand with the classical world. Among these three, wave-particle duality epitomizes the fundamental nature of quantum mechanics. According to Bohr’s complementarity principle, quantum objects can behave as a particle as well as a wave depending on the measurement apparatus. Feynman reckoned this wave-particle duality as the fundamental mystery of quantum mechanics\cite{feynman}. The reason for this counter-intuitive nature is unknown to us. Many attempts in the form of delayed-choice gedanken experiments\cite{Lref1,whdq3,othr1,othr2} have been made to find out when quantum objects behave as wave or particle. An experiment highlighting the inadequacy of the complementarity principle in its usual form is also proposed.\cite{entg1} Different experimental realizations of ``quantum-eraser" were studied to investigate and to control the exact nature of quantum objects.\cite{Lref3,Lref4,Lref2,Lref5}. 
In an alternative model of Young's double-slit experiment, \cite{whd,whd1}, Wheeler also attempted to find the solution to this mystery. Hidden-variable (HV) theories can also model the results of quantum mechanics, making it an alternative description of quantum theory. HV theories, based on wave-particle objectivity, determinism, and local independence was studied, hoping that it could reproduce the outcomes of quantum mechanics\cite{entg2} but if we take into account all the three criteria mentioned above, even the HV theories fail to reproduce the quantum mechanical predictions. With the advent of quantum technologies, one can test experimentally the HV theories and quantum mechanical predictions. The investigation of wave-particle duality paves a way to differentiate between these two theories.\\

\begin{figure}
	\centering
	\includegraphics[scale=0.4]{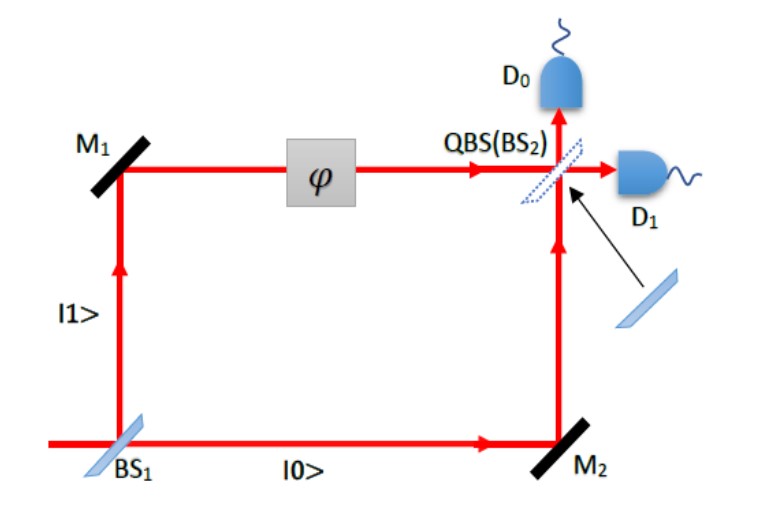}
	\caption{\textbf{Quantum controlled Wheeler's delayed-choice experiment:} $BS_1$ and $BS_2$ are beam splitters inside Mach-Zehnder interferometer (MZI). $\phi$ is phase shifter, $D_0$ and $D_1$ are detectors. When $BS_2$ is present, the MZI is in `closed' configuration and when $BS_2$ is absent, the MZI is in `open’ configuration. QBS represents a beam splitter which is controlled by a two-level quantum system in superposition.}\label{fig1}
\end{figure}

The wave-particle duality of quantum objects can be illustrated by delayed-choice experiments in a Mach-Zehnder interferometer (MZI) as depicted in Fig.\ref{fig1}. In the single-photon delayed-choice experiment, a photon is directed to a Mach-Zehnder interferometer one by one where it splits through a beam-splitter $BS_1$into two paths, represented by $\ket{0}$ and $\ket{1}$. A phase-shifter in one of the paths introduces a relative phase $\phi$ shift between the two paths. In the `closed' setup configuration the two paths are recombined by a second beam-splitter $BS_2$ thus the which-way information (path information) of the photon is erased and the photon exhibit an interference pattern recorded by each detector $D_0$ and $D_1$ where the intensities oscillate with relative phase difference $\phi$ between two paths, reveals the wave nature of the photon. In the `open' setup configuration, second beam splitter $BS_2$ is removed and each detector behind the two paths $D_0$ and $D_1$ clicks at a time with an equal probability of $\frac{1}{2}$, revealing particle property of the photon.\\

Jacques et. al. \cite{whdc} have demonstrated a delayed-choice experimental setup where the introduction of the second beam-splitter was randomly controlled by a random number generator (RNG) after the photon has already passed through the first beam-splitter $BS_1$ to avoid any causal connection between the selection of the paths and the presence of second beam-splitter $BS_2$. In a more recent development Ionicioiu and Terno \cite{whdq} have come up with the new idea of the quantum delayed-choice experiment where the introduction of the second beam-splitter $BS_2$ is controlled by the superposition state of a two-level quantum system. So the presence and absence of the beam-splitter is decided in a quantum way. The results not only describe the intrinsic wave-particle duality of quantum objects but also show the morphing behavior between the wave and the particle property \cite{whdq1,whdq2}. This morphing of wave and particle nature is the evidence that the quantum mechanical predictions about the nature of a photon should show a contradiction with the prediction of any local hidden variable model which attempt to designate the intrinsic wave or particle characteristic to the photons even before the final measurement.\\

The availability of open-source quantum computer via the ``IBM quantum experience" cloud-based platform gives the provision to check different fundamental features of quantum physics \cite{fun1,fun2,fun3,fun4,fun5,fun6,fun7}. Using this platform we can realize real quantum experiments and simulations. In this article, we utilize the IBM’s superconducting chip to demonstrate and perform two different extensions of the delayed-choice experiment. We experimentally demonstrate the quantum delayed choice scheme and investigate the variation of probabilities (which is equivalent to the interference intensities in real photonic setup), observing a continuous transition between wave and particle behaviour. By considering the entanglement assisted quantum delayed-choice scheme we show that the predictions of a local hidden variable model significantly differ from the quantum mechanical predictions and fail to explain the results of the experiment.

\section*{Theory}
\begin{figure}
	\centering
	\includegraphics[scale=0.25]{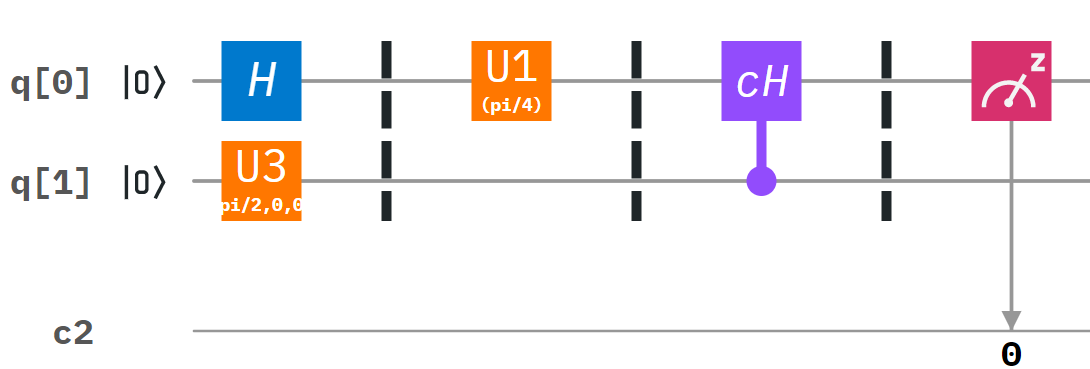}
	\caption{\textbf{Equivalent quantum circuit for quantum delayed-choice experiment:} The system qubit q[0] is initially prepared in $\Ket{0}$. $H$ is the Hadamard operator. $U_1$ represents the phase shift operator. The ancilla qubit q[1] is prepared in the superposition state by $Y_\alpha = e^{i\alpha\sigma_y}$ acting on the $\Ket{0}$ which essentially controls the presence or absence of second Hadamard gate (controled-Hadamard gate).}\label{fig2}
\end{figure}

In the following, we briefly explain the basic idea of quantum delayed choice experiment. Consider a photon that is inserted into the Mach-Zehnder interferometer as described in Fig.\ref{fig1}. For our experiment, we model an equivalent quantum circuit for the above photonic setup in the IBM quantum experience platform using proper gates which is shown in Fig.\ref{fig2}. It includes the Hadamard operator $H$ which performs the function of the first beam-splitter $BS_1$  and a phase shift operator $U1$ which acts as a phase shifter $\phi$. Control-Hadamard operator does the job for the second beam splitter $BS_2$ whose presence or absence is determined in a quantum way (quantum beam-splitter). Measurement in computational basis is performed at the very end of the quantum circuit. There are two ways in which quantum delayed-choice experiments can be performed. One is by using only a single ancilla while in the other approach one can use an entangled ancilla given by the EPR pair. Here we have used the former one to show the wave-particle morphing behavior while the latter is used to describe the incompatibility between quantum theory and a local Hidden variable model. \\

\textit{\textbf{With Single ancilla:}} q[0] and q[1] are the system qubit and ancilla qubit respectively. The preparation of ancilla qubit is done by rotating the initial state $\ket{0}$ by an angle $\alpha$ around y-axis using operator $Y_\alpha = e^{i\alpha\sigma_y}$. $\ket{0}$ and $\ket{1}$ states of the ancilla correspond to the absence and presence of the second Hadamard gate in the system qubit q[0] respectively. The insertion $(\alpha=\pi/2)$ of the second Hadamard gate results in the detection probability of any particular final state ($\ket{0}$ or $\ket{1}$) of the system qubit to vary sinusoidally with a relative phase shift $\phi$, giving rise to a phenomenon equivalent to interference. From this, it can be concluded that the quantum system behaves like a wave. On the other hand, the removal $(\alpha=0)$ of the second Hadamard gate shows that the probability of obtaining a final state ($\ket{0}$ or $\ket{1}$) of the system qubit is a constant $(1/2)$ concerning any change in phase shift $\phi$, giving strong evidence of particle-like nature of the system qubit. In our investigation, we have put our ancilla qubit in a superposition of quantum states ${\cos{\alpha}\ket{0}+\sin{\alpha}\Ket{1}}$ so that morphing of wave and particle nature can be observed as the values of $\alpha$ vary from 0 to $\pi/2$.\\

Initially the system qubit q[0] is in $\Ket{0}$ state. The Hadamard operation will result in the superposition state  $\frac{\Ket{0}+\Ket{1}}{\sqrt{2}}$. This state is further acted upon by an $U_1$ gate which will introduce variable phase shift to the state $\Ket{1}$, so the state transforms to $\frac{\Ket{0}+ e^{i\phi}\Ket{1}}{\sqrt{2}}$. $P_0= \Ket{0}\bra{0}$ and $P_1= \Ket{1}\bra{1}$ act as two orthogonal projectors onto the eigenstates of the system. Now, when there is no second Hadamard gate present in the system qubit, the final state remains $\Ket{\psi_p} = \frac{\Ket{0}+ e^{i\phi}\Ket{1}}{\sqrt{2}}$ . So the probabilities of detecting the system in state $\Ket{0}$ and $\Ket{1}$ are given by the expectation values of the projectors respectively as,
\begin{align}
E_{p_0} = \bra{\psi_p}P_0\Ket{\psi_p} = \frac{1}{2}\nonumber \\
E_{p_1} = \bra{\psi_p}P_1\Ket{\psi_p} = \frac{1}{2}
\end{align}
The above equation suggests that $E_{p_0}$ and $E_{p_1}$ do not depend on the phase shift $\phi$, which is a clear indication of particle like behaviour of the system. So it indicates that no interference will be observed in the real experimental setup and the visibility $\frac{E_{max} - E_{min}}{E_{max} + E_{min}}$ of interference will be zero. But when the second Hadamard gate is present in the circuit the final state becomes $\psi_w = \cos{\frac{\phi}{2}}\Ket{0} + sin{\frac{\phi}{2}}\Ket{1} $ and the corresponding probability of detecting the state $\Ket{0}$ and $\Ket{1}$ is 

\begin{align}
E_{w_0} = \bra{\psi_w}P_0\Ket{\psi_w} = \cos^2{\frac{\phi}{2}}\nonumber \\
E_{w_1} = \bra{\psi_w}P_1\Ket{\psi_w} = \sin^2{\frac{\phi}{2}}
\end{align}

which indicates the wave-like behavior of the system.

Till now we have considered a definite state of the ancilla qubit and so does the presence or absence of the second Hadamard gate. But when we prepare the ancilla qubit in a superposition state, namely to ${\cos{\alpha}\ket{0}+\sin{\alpha}\Ket{1}}$ by operating a rotation $e^{i\alpha\sigma_y}$ to the ancilla qubit then the system qubit and the ancilla qubit become entangled and the combined state can be written as\\
\begin{equation}
\Ket{\psi_f}= \cos\alpha \Ket{\psi_p}\Ket{0}_a + \sin\alpha \Ket{\psi_w}\Ket{1}_a
\end{equation}
where the subscript `a' corresponds to the ancilla qubit.
The reduced density operator for the system is obtained by tracing out the ancilla
\begin{equation}
\rho_{s}= \cos^2{\alpha} \Ket{\psi_p}\bra{\psi_p} + \sin^2{\alpha} \Ket{\psi_w}\bra{\psi_w}
\end{equation}
where $\rho_{s}$ is equal to the $Tr_{a}[\rho_{f}$].
The intensity in each detector (equivalent to two distinct projector $D_0$ and $D_1$) can be obtained from the expectation values respectively as
\begin{align}
E_{s_0} = Tr[\rho_s D_0]=\frac{Cos^2{\alpha}}{2}+\sin^2{\alpha}\cos^2{\frac{\phi}{2}}\nonumber\\
E_{s_1} = Tr[\rho_s D_1]=\frac{Cos^2{\alpha}}{2}+\sin^2{\alpha}\sin^2{\frac{\phi}{2}}
\label{eq:5}
\end{align}
where, $D_0 = \Ket{0}\bra{0}$ and $D_1 = \Ket{1}\bra{1}$. It is clear from the above equation that when $\alpha=0$ the probability of obtaining the state of system qubit in $\Ket{0}$ or $\Ket{1}$ is $\frac{1}{2}$ whereas for $\alpha= \pi/2$ we will get a sinusoidal variation of the probability (with respect to change in $\phi$). So from the above theoretical analysis we can infer that in the first case there will be no change in the detection probability with respect to the change in $\phi$, giving rise to the particle like nature of the system which is in contrast with the later case, where the detection probability will show phase dependent sinusoidal interference like behavior, which is a signature of the wave property.\\

\begin{figure}
	\centering
	\includegraphics[width=0.8\linewidth]{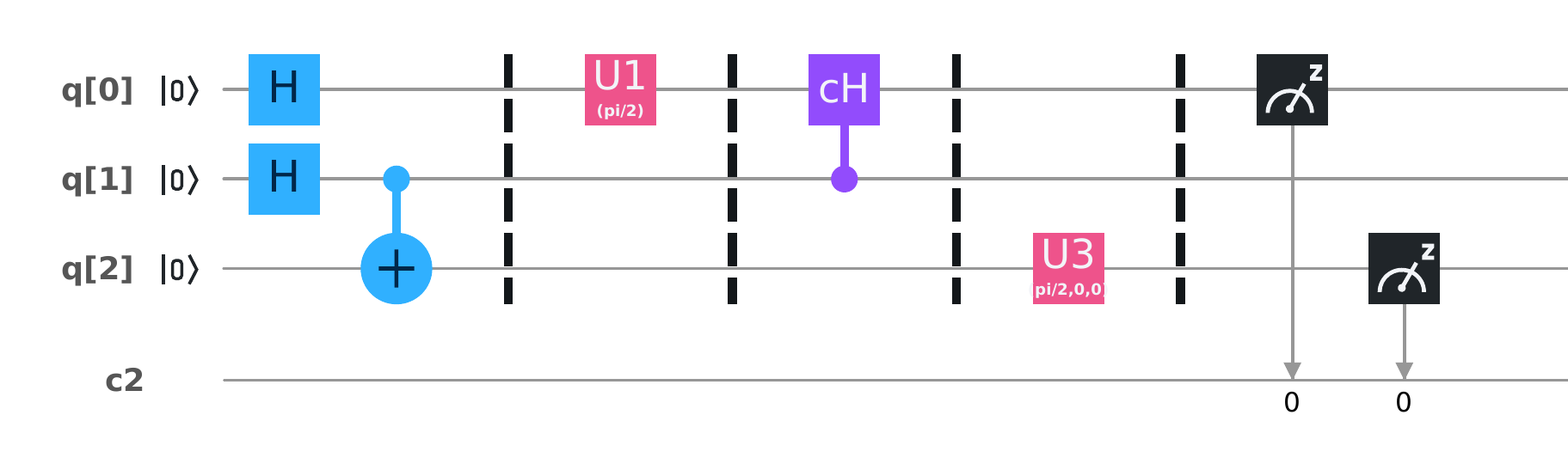}
	\caption{\textbf{Equivalent quantum circuit for entanglement assisted delayed-choice experiment:} q[0] is the system qubit and q[1] and q[2] are the entangled ancilla qubits. The Hadamard gate and CNOT gate are used for generating an EPR pair. The rest of the configuration is similar to the circuit in Fig \ref{fig2}. }\label{fig3}
\end{figure}

\textit{\textbf{With Entangled ancilla:}} This experimental scheme was first proposed by Ionicioiu et. al. \cite{entg2} and later performed by Xin et. al. \cite{entg3} in NMR platform. The equivalent quantum circuit for this experiment is shown in Fig.\ref{fig3} where we use an EPR pair (q[1] and q[2]) as ancilla qubits. This particular scheme provides a way of testing the predictions of quantum theory and HV models. The main advantage of an entangled ancilla qubit over a single qubit is that it not only gives the quantum control on the Hadamard gate but we can also take the advantage of space-like separated events. Taking this into account we introduce a rotation to q[2] before measuring it in a computational basis. 

As shown in Fig.\ref{fig3}, first the ancilla qubits q[1] and q[2] are made entangled by using Hadamard gate and a controlled not gate so that initial state of the ancilla becomes $\frac{\Ket{00}+\Ket{11}}{\sqrt{2}}$. System qubit q[0] is initially acted upon by a Hadamard gate which results in a superposition $\frac{\Ket{0}+\Ket{1}}{\sqrt{2}}$ while $U_1$ produces a relative phase change $\phi$ between $\Ket{0}$ and $\Ket{1}$. Now q[1] acts as a control qubit which put the restriction on the second Hadamard gate in q[0]. In other word, the state of the second qubit is viable for the emergence of two different nature of system qubit. Since q[2] is entangled with the q[1], we have the freedom of giving ancilla a rotation  $e^{i\alpha\sigma_{y}}$, even after q[0] and q[1] have already interacted with each other. Finally, composition of all the above actions produces an output state given by: 

\begin{equation}
\begin{split}
\Ket{\psi} & = \frac{ie^{i\phi}}{\sqrt{2}}[ (\cos{\alpha}\Ket{\psi_p}\Ket{0}+\sin{\alpha}\Ket{\psi_w}\Ket{1})_{q_0q_1}\Ket{0}_{q_2}\\
& \hspace{1cm} -(\sin{\alpha}\Ket{\psi_p}\Ket{0}-\cos{\alpha}\Ket{\psi_w}\Ket{1})_{q_0q_1}\Ket{1}_{q_2}]
\end{split}
\end{equation} 

where, $\Ket{\psi_p}=\frac{\Ket{0}+e^{i\phi}\Ket{1}}{\sqrt{2}}$ and $\Ket{\psi_w}=e^{i\frac{\phi}{2}}[cos(\frac{\phi}{2})\Ket{0}- isin(\frac{\phi}{2})\Ket{1}]$.
The reduced density matrix for the system is evaluated by tracing out ancilla qubits and is given by, $\rho_A =Tr_{AB}[\rho]=Tr_{AB}[\Ket{\psi}\bra{\psi}]$. With this reduced density matrix the intensity in one of the detector say $D_0$ is calculated as: $E_A = Tr[\rho_A\Ket{0}\bra{0}]$. Now as we are making a measurement in third qubit, two expressions for intensities arise for two different post-selections($\Ket{0} and \Ket{1}$) of the third qubit and are given by:
\begin{align}
E_{A_0} = \cos^2{\frac{\alpha}{4}}+\frac{\sin^2{\alpha}\cos^2{\phi/2}}{2} \nonumber \\
E_{A_1} = \sin^2{\frac{\alpha}{4}}+\frac{\cos^2{\alpha}\cos^2{\phi/2}}{2} \label{eq:7}
\end{align}
The above equations well explain the quantum mechanical predictions about the intensities for this particular configuration. Considerable departure is observed when we start formulating statistics for the same through a local HV model. If any HV model is compatible with quantum mechanical predictions then it should reproduce the quantum probability distribution $P(q_0,q_1,q_2)$ when summed over all the Hidden variables, namely:
\begin{equation}
P(q[0],q[1],q[2])=\sum_{\lambda} C(q[0],q[1],q[2],\lambda) \label{eq:8}
\end{equation}
Where, $C(q[0],q[1],q[2],\lambda)$ denotes the probability distribution due a hidden variable satisfying three classical assumptions given by:\\

{\textbf{1. Wave particle objectivity:}} According to the wave particle objectivity the nature of a photon is attributed as two distinct characteristics with no overlap. So with respect to an open (q[1]=0) Mach-Zehnder interferometer setup shown in Fig.\ref{fig2}, the conditional probability at any of the detector $D_0$($D_1$) is statistically represented as,
\begin{equation}
P(q[0] | q[1]=0,\lambda)= (\frac{1}{2},\frac{1}{2}) \hspace{1cm}\forall\hspace{0.25cm} \lambda\in S_p
\end{equation}
while for a closed (q[0]=1) one it reads, 
\begin{equation}
P(q[0] | q[1]=1,\lambda)=(\cos^2{\frac{\phi}{2}},\sin^2{\frac{\phi}{2}}) \hspace{1cm}\forall\hspace{0.25cm} \lambda\in S_w
\end{equation}
Simply, $S_p \cap S_w =\Phi$, where $\Phi$ is a null set. Otherwise there will be elements for which wave particle duality occurs.\\

{\textbf{2.Determinism:}} The theory of determinism states that the outcomes of an event should have a preexisting value which is independent of the act of measurement. From equation \eqref{eq:8} it is seen that whatever value $q_i$' possess, probabilities are given by summing over some preexisting values of $\lambda$.\\

{\textbf{3.Local independence:}} Classically at the level of outcomes two spatially separated systems always possess locally independent hidden variables. So the joint probabilities for any event are mutually exclusive and can be represented as, 
\begin{equation}
P(\lambda)=P_1(\lambda_1)P_2(\lambda_2)
\end{equation}
where $\lambda_1$ and $\lambda_2$ are hidden variables for two spatially separated systems. In the current case qubit, $q_1$ and $q_2$ represents such spatially separated systems which are acted upon by two different hidden variables which are mutually exclusive in an outcome level theory.\\

Taking into account the aforementioned assumptions the same intensities takes the following form \cite{entg2}
\begin{equation}
E_{A_1}= E_{A_0} = \frac{1}{4}+\frac{\cos^2{\phi/2}}{2}\label{eq:12}
\end{equation}
which different from the predictions of quantum theory. Though from equation\eqref{eq:7} and equation\eqref{eq:12} it seems that the nature of the variation of intensity is same for both of the cases but their magnitude differs significantly for values $\cos{2\alpha}\neq 0$. In the next section we will verify and discuss all our theoretical predictions and experimental outcomes obtained from IBM's superconducting quantum processor.
\begin{figure}
	\centering
	\includegraphics[scale=0.5]{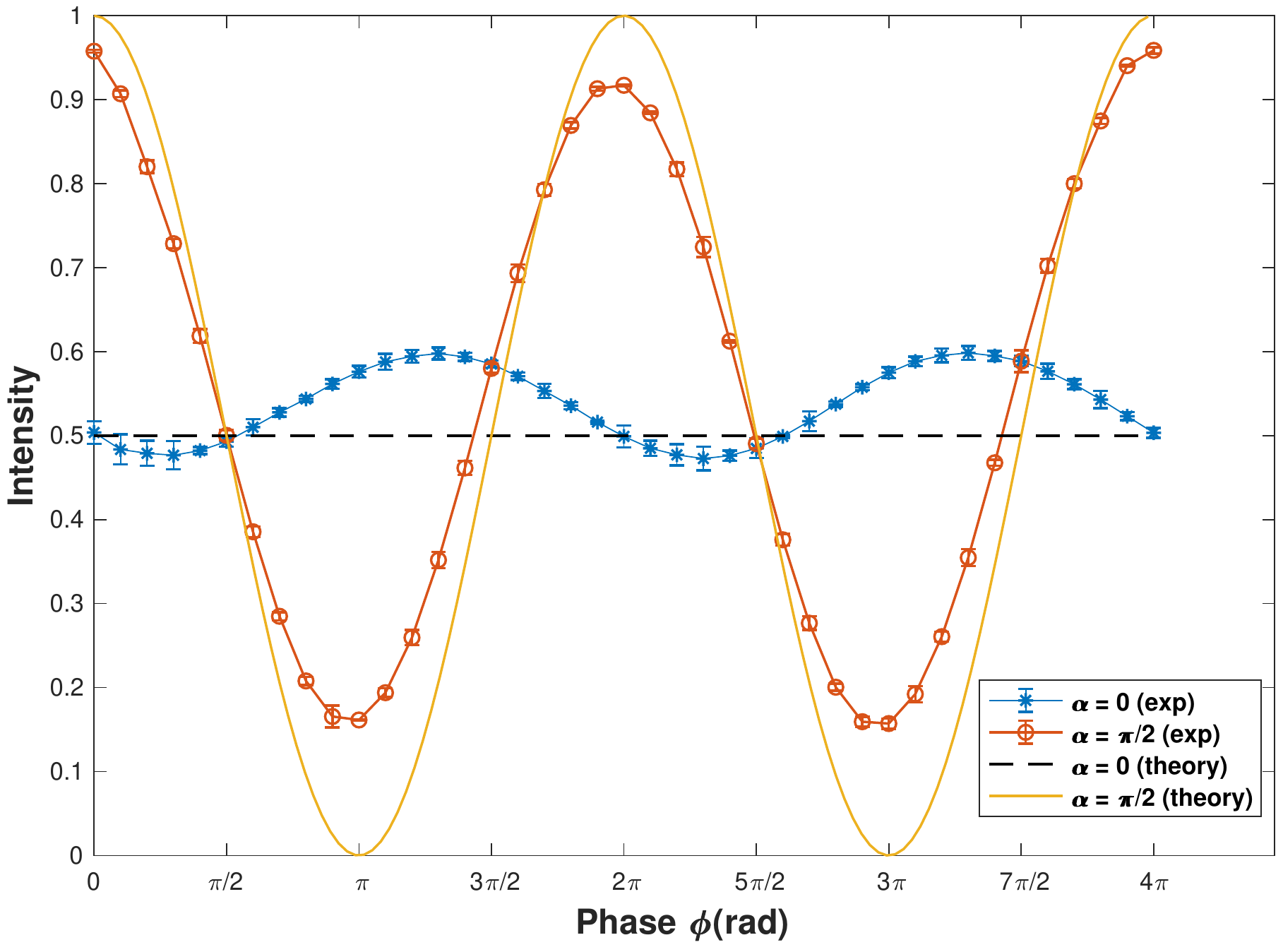}
	\caption{\textbf{Experimental results for the quantum delayed-choice experiment:} Variation of intensity with phase change $\phi$ for two extreme values of $\alpha = 0$ (red) and $\alpha = \pi/2 $ (blue) is shown. Points with error bars represent the experimental values of intensity variation with phase shift $\phi$ while the solid lines describes the theoretical predictions. The measured data shows good agreement with theoretical predictions.
}\label{fig4}
\end{figure}

\section*{Experimental Analysis and Discussion} 
To investigate the quantum delayed choice experiment, we have used IBM's 5 qubit quantum processor `ibmq\_5\_tenerife' and we have implemented the quantum circuit as described in FIG. \ref{fig1}. To obtain the intensity distribution, first, we measure the probability of obtaining the output state $\Ket{0}$ for a given value of the rotation angle $\alpha$ and then for each value of $\alpha$ we vary the relative phase $\phi$. We analyze the above situation for two extreme cases: $\alpha=0$ and $\alpha=\frac{\pi}{2}$ respectively. For $\alpha=0$ we have nearly constant intensity while for $\alpha=\frac{\pi}{2}$ the intensity is varying sinusoidally with $\phi$ as shown in Fig.\ref{fig4}. The most interesting case occurs when we investigate the nature of intensity for the values of $\alpha$ lying in the range between 0 to $\pi/2$. It is when the Hadamard gate is in a state of superposition of being present and absent.
\begin{figure}[]
    \centering
    \subfloat[]{\includegraphics[width=0.5\linewidth]{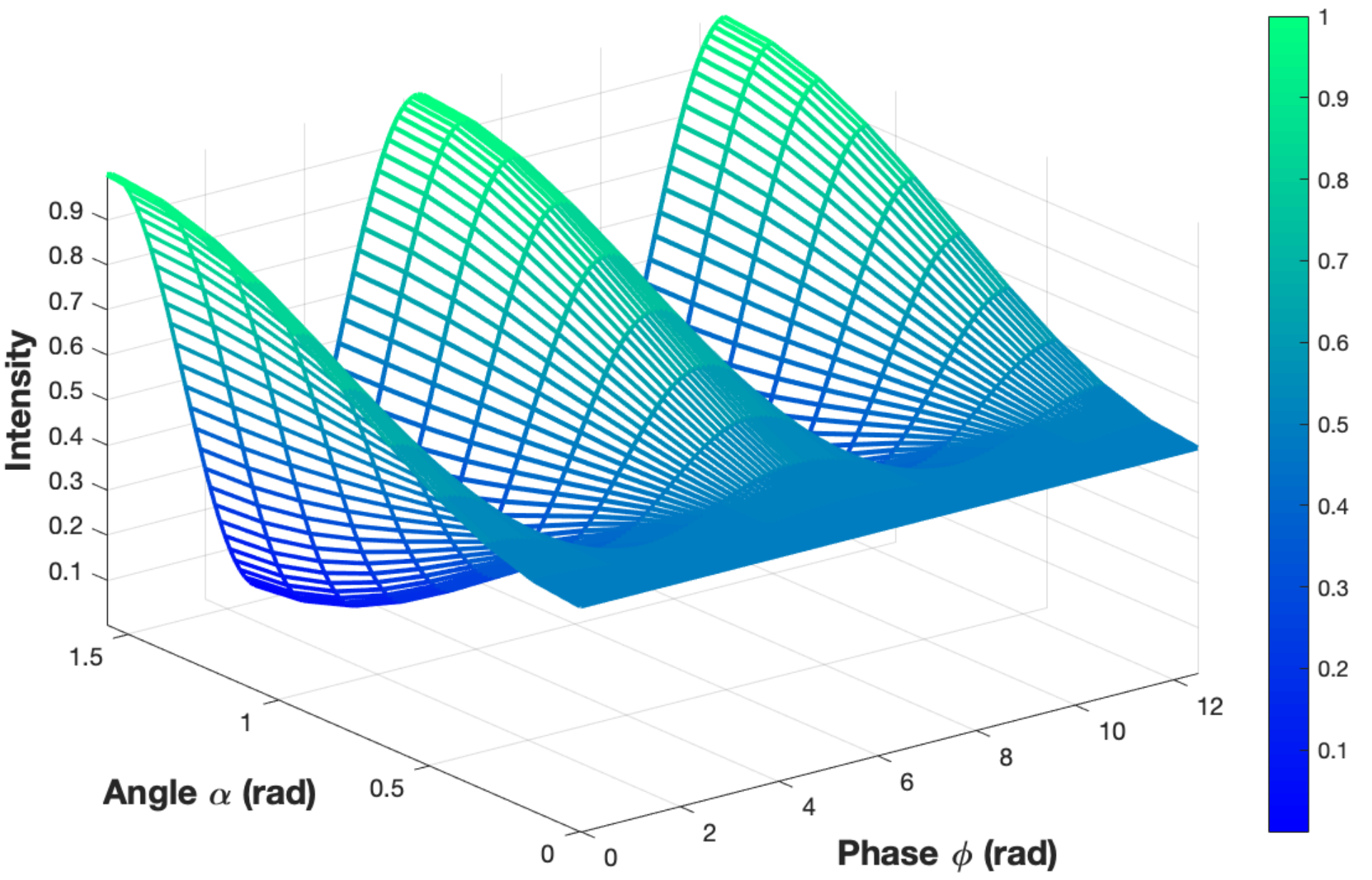}}
    \subfloat[]{\includegraphics[width=0.5\linewidth]{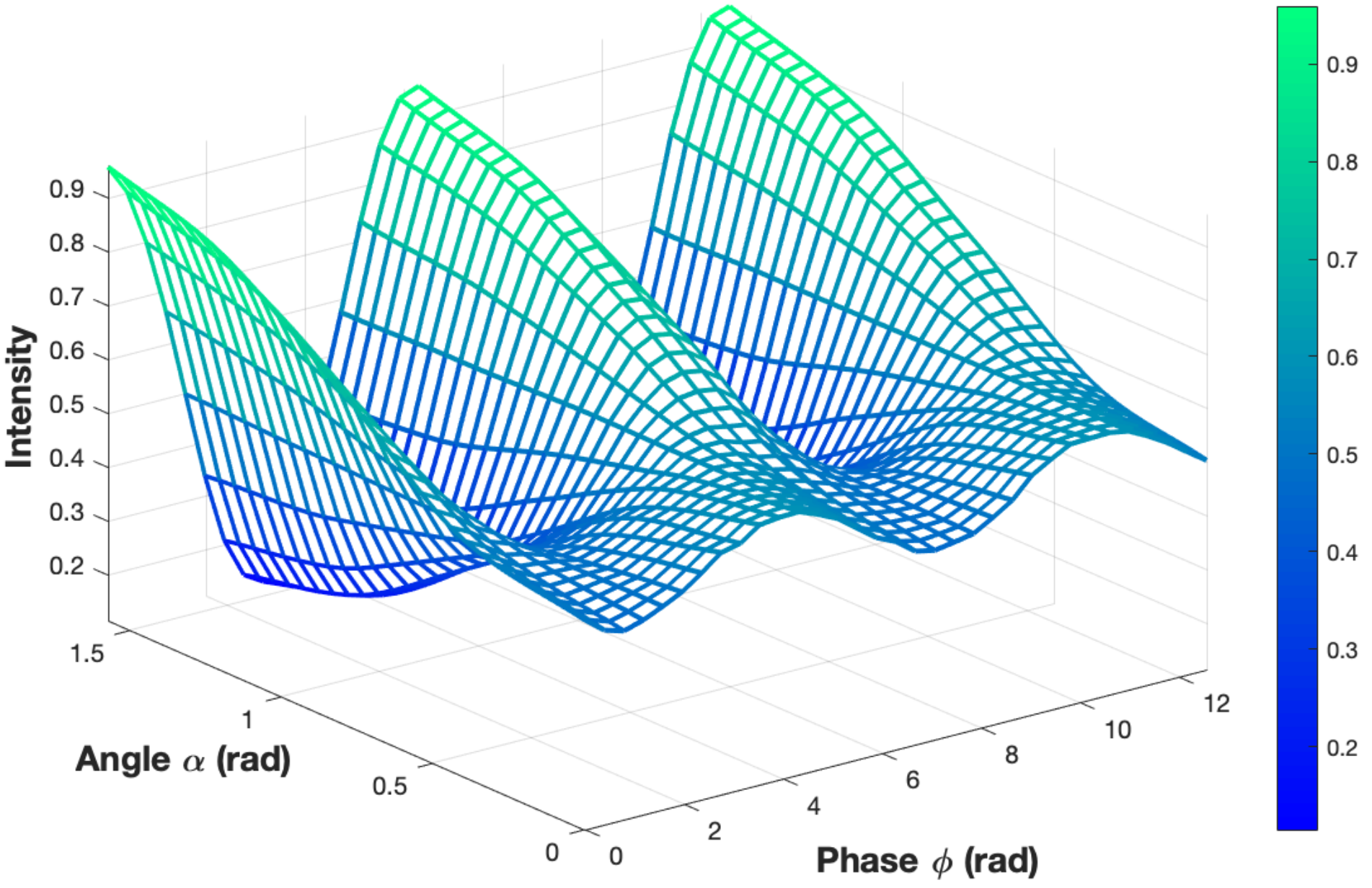}}
    \caption{\textbf{Characterization of wave-particle morphing behaviour:} 3-D graphs are plotted for intensity distribution as a function of phase difference $\phi$ for different values of rotation angle $\alpha$. The wave particle morphing behavior can be observed for intermediate values of $\alpha$ between $0$ to $\frac{\pi}{2}$. a) Shows the simulated intensity profile and b) shows the plot for measured experimental data. The observed data shows good agreement with theoretical predictions (see equation \eqref{eq:5}).
 }\label{fig5}
\end{figure}
For this range we observe a continuous transition between particle and wave nature of the system qubit as shown in Fig.\ref{fig5}.
\begin{figure}[]
    \centering
    \subfloat[]{\includegraphics[width=0.4\linewidth]{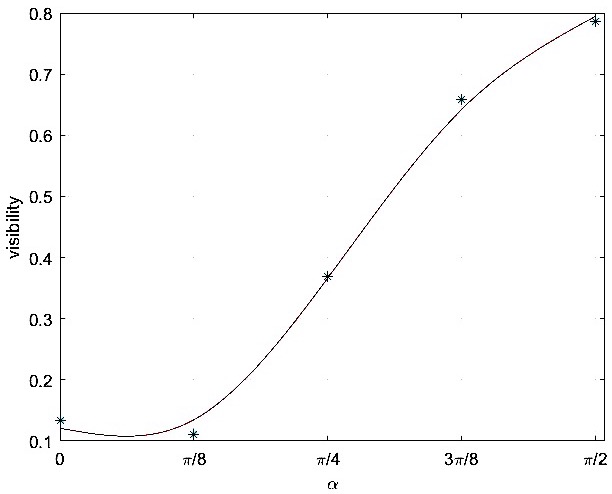}}
    \subfloat[]{\includegraphics[width=0.4\linewidth]{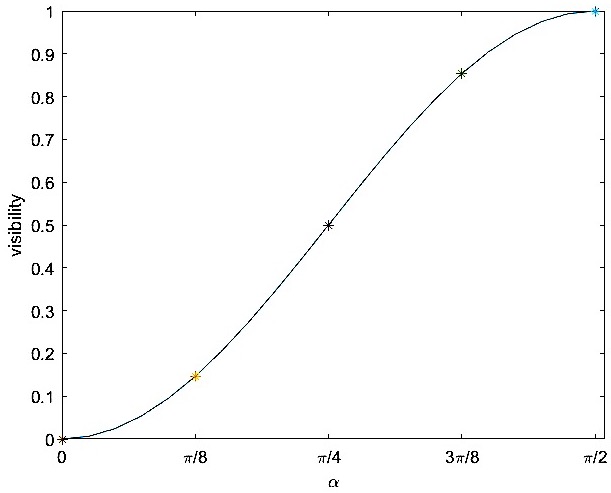}}
    \caption{\textbf{Visibility curve for verifying wave-particle morphing behavior:} The variation of visibility
    $\frac{E_{max}-E_{min}}{E_{max}+E_{min}}$ with rotation angle $\alpha$ are plotted. a) shows the experimental plot and b) describes the theoretical plot of visibilities. It can be seen that the experimental data matches well with the theoretical predictions.  } \label{fig6}
\end{figure}

For each value of rotation angle $\alpha$, the maximum and minimum value of intensity is noted and from that visibility, the graph is obtained which is shown in Fig.\ref{fig6}. From the graph, we can see that for $\alpha=0$, visibility is nearly zero though it should be exactly zero according to the theoretical prediction. This small error can be characterized by individual gate errors but the overall variation of visibility with $\alpha$ is similar as given by theoretical one. A null value of visibility corresponds absence of any interference but as $\alpha$ increases the value of visibility also gradually increases causing observable interference. So our analysis demonstrates how wave-particle morphing is coming out from the uncertainty in the quantum control experimental setups and how it is different from a classical delayed choice experiment.\\

\begin{figure}[]
    \centering
    \includegraphics[scale=0.65]{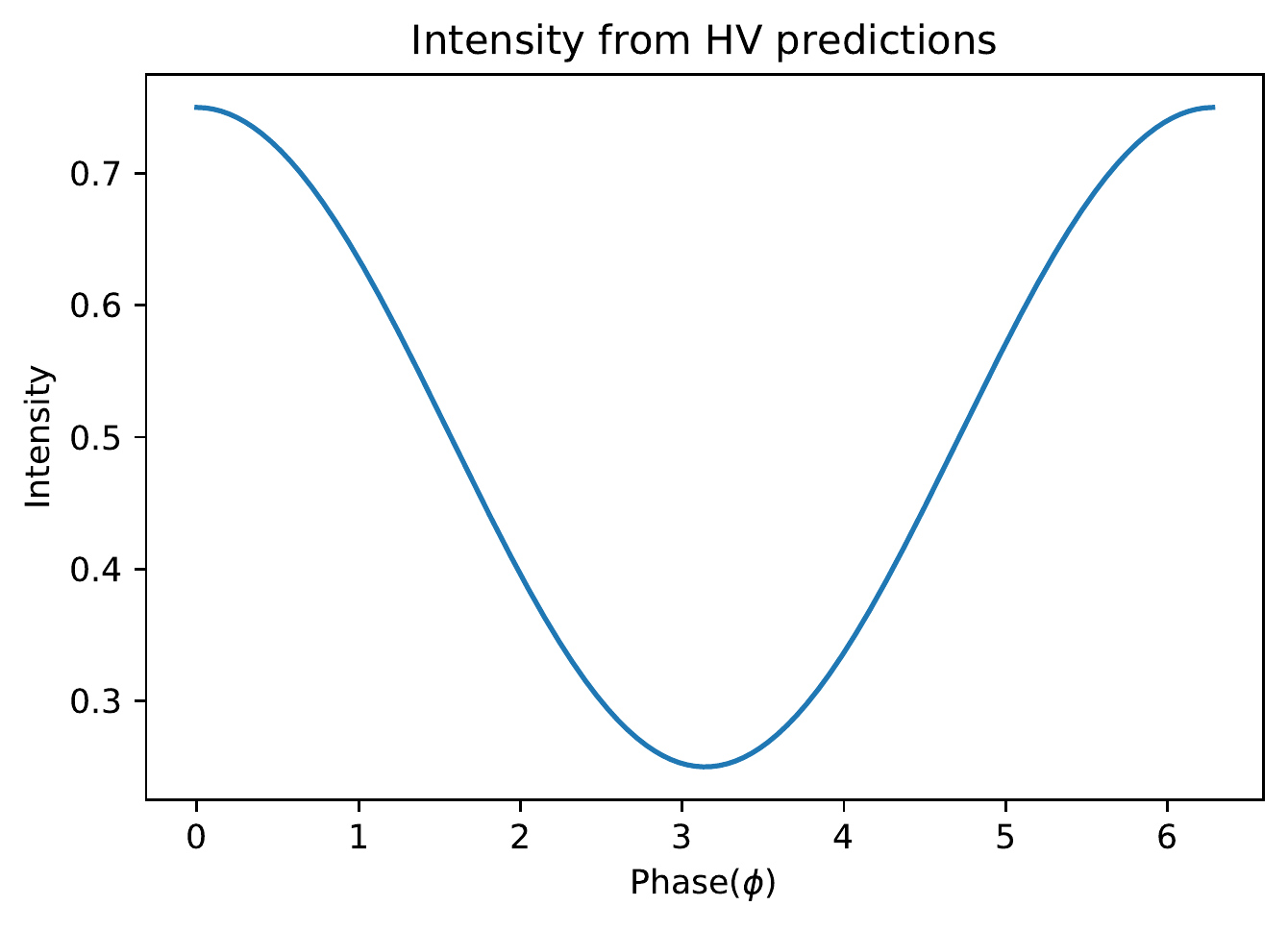}
    \caption{\textbf{Verification of wave-particle morphing behavior:} Variation of intensity vs phase shift $\phi$ (ranging from 0 to 2$\pi$), from HV predictions. Since,  for both post selected states $\Ket{0}$ and $\Ket{1}$,  the intensities are same, only one graph is shown.}\label{fig7}
\end{figure}

\begin{figure}[]
    \centering
    \includegraphics{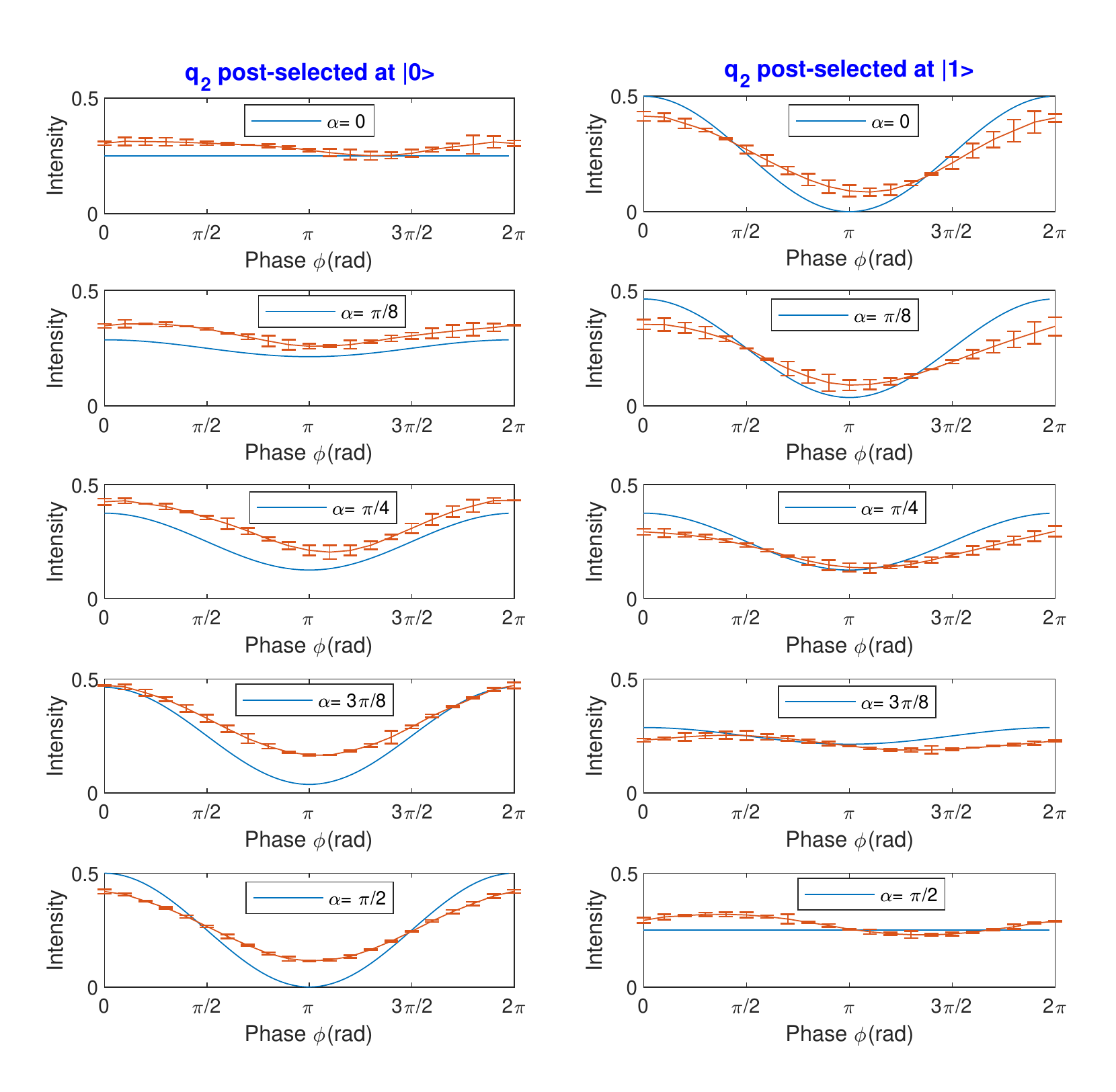}
    \caption{\textbf{Experimental Intensities:} Variation of  intensities versus phase difference $\phi$ for different values of $\alpha$ . The solid blue lines show the quantum mechanical predictions whereas points with error bars represent the experimental values. This is the data resulting from the circuit shown in Fig. \ref{fig3}, the experiment was run for $\phi$ ranging from 0 to 2$\pi$ on a real quantum device. This experiment was run three times for statistical accuracy. }\label{fig8}
\end{figure}

To realize the entanglement assisted delayed choice experiment, we have implemented the equivalent quantum circuit shown in FIG \ref{fig3} in IBM's 14 qubits quantum processor `ibmq\_16\_melbourne'. The experimental parameters of the device are presented in Table \ref{qgs_tab1} where the coherence time, relaxation time, gate error and readout error are respectively denoted by $T_{1}$, $T_{2}$, GE and RE. The superconducting qubits q[8], q[9] and q[10] are used for this experiment. For consistency we have denoted q[8] = q[0], q[9] = q[1] and q[10] = q[2] respectively in Fig \ref{fig3}. First q[1] and q[2] are made entangled through a Hadamard and a CNOT gate. A rotation in q[2] is introduced before the measurement through $U_3$ gate which is introducing a dependency of the intensity on $\alpha$. We can wisely choose the rotation of $\alpha$ in q[2] after the interaction of the first two qubits to take advantage of space-like separated events in a quantum computer. The variation of intensity as predicted by the HV theory (see equation \eqref{eq:12}) is shown in Fig.\ref{fig7}. It is to be noted from equation \eqref{eq:12} that the intensity is not a function of $\alpha$ so for HV prediction, the variation of intensity is the same for both post selected states $\Ket{0}$ and $\Ket{1}$ while quantum mechanical prediction of intensity depends on $\alpha$ through equation \eqref{eq:7}. We have plotted variation of intensities against phase shift $\phi$ for different values of alpha and is shown in Fig.\ref{fig8}

The underpinning reason behind the incompatibility between QM and HV predictions is that while in a local HV theory, objective reality exists, in the case of quantum mechanics it is not true. So morphing between wave and particle nature is seen as a part of our experiment. This morphing is due to the fact that in the case of QM predictions, intensity depends on $\alpha$ and as it changes, the magnitude of intensity changes even for the same phase $\phi$. On the other hand, since HV prediction of intensity is independent of $\alpha$ the intensity remains the same for particular $\phi$ even though the rotation angle $\alpha$ changes. Our experiment demonstrates the fact that wave-particle morphing is a purely quantum phenomenon that can not be reproduced from the HV theory consisting of wave-particle objectivity, determinism, and local independence.

\begin{table}[]
\centering
\begin{tabular}{ c c c c c c }
\hline
\hline
Qubits &  $T^{\dagger}_{1}$ ($\mu s$) & $T^{\ddagger}_{2}$ ($\mu s$) & GE$^{||}$)  & RE$^{\perp}$ & CNOT Error\\
\hline
q[8] & 64.225 & 94.455 & 0.004 & 0.044 & CX9\_8: 0.06 \\
q[9] & 37.253 & 62.463 & 0.005 & 0.044 & -\\
q[10] & 66.177 & 71.602 & 0.003 & 0.039 & CX9\_10: 0.05 \\ 
\hline
\hline
\end{tabular}\\
$\dagger$ Coherence time, $\ddagger$ Relaxation time, $||$ Gate Error, $\perp$ Readout Error. 
\caption{\textbf{The device parameters of the quantum processor `ibmq\_16\_melbourne'.}}
\label{qgs_tab1}
\end{table} 

\section*{summary}
In conclusion, we have studied the delayed-choice scheme to investigate the wave-particle duality of quantum objects. In this regard, IBM's online cloud-based quantum computer provides a perfect platform for investigating such phenomena. In this article, we have successfully demonstrated two different extensions of a delayed-choice experiment namely quantum delayed-choice experiment where we have used quantum controlled Hadamard gate to tune the wave and particle behavior of a quantum system and entanglement-assisted quantum delayed-choice experiment to show when all three classical assumptions of HV is combined, it leads to conflict with the quantum theory. In our results, we found a good agreement between the experimentally obtained intensities, visibilities and continuous morphing of wave and particle behavior with the predictions of quantum theory.

\section*{Acknowledgments}
The authors acknowledge the use of IBM Quantum Experience platform for this work and grateful to IBM team. P.D.C. and S.M. and A.A.B. and N.N.H acknowledge the help and hospitality of IISER kolkata during the project. The discussions and opinions developed in this paper are only those of the authors and do not reflect the opinions of IBM or IBM Q experience team.

\end{document}